\documentclass[aps,prd,showpacs]{revtex4-1}
\usepackage{amsmath,amssymb}
\usepackage{indentfirst,latexsym,bm}
\begin{document}
\title{Extracting Rephase-invariant $CP$ and $CPT$ Violating Parameters from Asymmetries of   Time-Ordered Integrated Rates of Correlated Decays of Entangled Mesons}
\author{Zhijie Huang}
\affiliation{Department of Physics, Fudan University, Shanghai, 200433, China}
\author{Yu Shi}
\email{yushi@fudan.edu.cn}
\affiliation{Department of Physics, Fudan University, Shanghai, 200433, China}
\begin{abstract}
We present a general model-independent formalism of measuring $CP$ and $CPT$ violating parameters through time-ordered integrated  rates of correlated decays of $C=\pm 1$ entangled states of neutral pseudoscalar mesons. We give the general formulae of   $CP$ and $CPT$ violating parameters in terms of four measurable asymmetries defined for the time-ordered integrated rates, applicable to all kinds of decay product. Two special cases which are often realized in experiments are discussed specifically.
\end{abstract}


Euro. Phys. J. C {\bf 72}, 1907 (2012)

\maketitle

\section{Introduction}

Precise measurement of $CP$ violating parameters in weak decays is an active topic. On the other hand, testing $CPT$ symmetry is also of great importance, as some physics beyond the standard model may lead to its breaking.  Nowadays, neutral $K$ and $B$ mesons are routinely produced as quantum-entangled or EPR-correlated pairs in $\phi$ and $B$ factories~\cite{domenico0,domenico,kloe2,old}. Hence it is highly interesting to explore their use  in examining $CP$ and $CPT$ symmetries~\cite{domenico0,domenico,kloe2,dunietz,buchanan,dambrosio,bernabeu,baji,shi}, and in measuring relevant parameters~\cite{soni,kittle}.  Especially, systematic proposals were made on exploring $CP$ and $CPT$ violating parameters in correlated decays of meson pairs in $C=-1$ state, by using decay rates and their asymmetries~\cite{buchanan,dambrosio}. In this paper, we  present a general and  rigorous calculation of time-ordered integrated rates of the correlated decays of $C=\pm1$ entangled meson pairs, taking into account both $CP$ violation and possible $CPT$ violation. Subsequently $CP$ and $CPT$ violating parameters are expressed as functions of the four asymmetries defined for the time-ordered integrated rates of the correlated decays of the two entangled states. Our result provides a general method of measuring rephase-invariant  $CP$ and  $CPT$ violating parameters in correlated decays of entangled mesons into any kind of product. This extends a rephase-invariant formalism of $CP$ and $CPT$ violating observables~\cite{3}  to  $C=\pm 1$  entangled mesons.

In Sec.~\ref{single}, we review the single-meson decays using two complex numbers parameterizing indirect $CP$ and $CPT$ violations. In Sec.~\ref{amplitude}, we consider the entangled states $|\psi_\pm\rangle$ and calculate the general expressions of the decay rates as functions of four single-meson decay amplitudes. In Sec.~\ref{integrated}, we integrate the decay rates over all times with two time orders, thereby obtain four time-ordered integrated rates. We give both exact results and approximate expressions up to the first order of the $CP$ and $CPT$ violating parameters. Subsequently, we obtaining the four real parameters as functions of the four asymmetries defined for the four  time-ordered integrated rates. In Sec.~\ref{special}, two special cases are treated for which the expressions are much simpler than the general expressions in Sec.~\ref{integrated}. Discussions and a summary are made in Sec.~\ref{summary}.

\section{Review of $CP$ and $CPT$ violating parameters in single-meson decays \label{single}}

An arbitrary state of a pseudoscalar meson can be written as
\begin{equation}
|M(t)\rangle = \psi(t)|M^0\rangle + \bar{\psi}(t) |\bar{M}^0\rangle,
\end{equation}
where $|M^0\rangle$ is the pseudoscalar meson state and $|\bar{M}^0\rangle$ is its antiparticle state, $\psi(t)$ and $\bar{\psi}(t)$ are superposition coefficients.
In Wigner-Weisskopf approximation, the time evolution of $|M(t)\rangle$ is described as
\begin{equation}
i\frac{d}{dt}{\psi(t) \choose \bar{\psi}(t)}=
\left( \begin{array}{cc}
H_{11} & H_{12} \\
H_{21} & H_{22}
\end{array}\right){\psi(t) \choose \bar{\psi}(t)},
\end{equation}
where $H_{ij}=M_{ij}-i{\Gamma_{ij}}/2$, with  the dispersive part $M_{ij}$ and  the absorptive part $\Gamma_{ij}$ are hermitian. Define~\cite{3}
\begin{eqnarray}
\delta_M&\equiv&\frac{H_{22}-H_{11}}{\sqrt{H_{12}H_{21}}},\\
\frac{q}{p}&\equiv&\sqrt{\frac{H_{21}}{H_{12}}} \equiv
\frac{1-\varepsilon_M}{1+\varepsilon_M}.
\end{eqnarray}
The  eigenstates of the Hamiltonian are
\begin{equation}
\begin{split}
|M_1\rangle=p_1|M^0\rangle+q_1|\bar{M}^0\rangle ,\\
|M_2\rangle=p_2|M^0\rangle-q_2|\bar{M}^0\rangle,
\end{split}
\end{equation}
with the respective eigenvalues
\begin{equation}
\begin{split}
\lambda_1=H_{11}+\sqrt{H_{12}H_{21}}(\frac{\delta_{M}}{2}+\sqrt{1+\frac{\delta_M^2}{4}}),\\
\lambda_2=H_{22}-\sqrt{H_{12}H_{21}}(\frac{\delta_{M}}{2}+
\sqrt{1+\frac{\delta_M^2}{4}}).
\end{split}
\end{equation}
Therefore we obtain
\begin{equation}
\begin{split}
|M^0\rangle=\frac{q_2|M_1\rangle+q_1|M_2\rangle}{p_1q_2+p_2q_1}, \\
|\bar{M}^0\rangle=\frac{p_2|M_1\rangle-p_1|M_2\rangle}{p_1q_2+p_2q_1}.
\end{split}
\end{equation}

For a single meson whose initial state is $|M^0\rangle$, we have
\begin{eqnarray}
|M^0(t)\rangle=\frac{p_1q_2e^{-i\lambda_1t}+p_2q_1e^{-i\lambda_2t}}{p_1q_2+p_2q_1}|M_0\rangle+\frac{q_1q_2(e^{-i\lambda_1t}-e^{-i\lambda_2t})}{p_1q_2+p_2q_1}|\bar{M}_0\rangle.
\label{m0}
\end{eqnarray}
Similarly, for a single meson whose initial state is $|\bar{M}^0\rangle$,
\begin{eqnarray}
|\bar{M}^0(t)\rangle=\frac{p_1p_2(e^{-i\lambda_1t}-e^{-i\lambda_2t})}{p_1q_2+p_2q_1}|M_0\rangle
+\frac{p_2q_1e^{-i\lambda_1t}+p_1q_2e^{-i\lambda_2t}}{p_1q_2+p_2q_1}|\bar{M}_0\rangle. \label{m6}
\end{eqnarray}

Define
\begin{equation}
\frac{1+\Delta_M}{1-\Delta_M}\equiv\frac{\delta_{M}}{2}+\sqrt{1+\frac{\delta_M^2}{4}},
\end{equation}
then~\cite{3}
\begin{eqnarray}
\frac{p_1}{q_1}=\frac{p}{q}\frac{1-\Delta_M}{1+\Delta_M}, \\
\frac{p_2}{q_2}=\frac{p}{q}\frac{1+\Delta_M}{1-\Delta_M}.
\label{piqifrac}
\end{eqnarray}
Therefore
\begin{eqnarray}
p_1&\propto&\frac{1}{\sqrt2}(1+\varepsilon_M)(1-\Delta_M), \\
q_1&\propto&\frac{1}{\sqrt2}(1-\varepsilon_M)(1+\Delta_M), \\
p_2&\propto&\frac{1}{\sqrt2}(1+\varepsilon_M)(1+\Delta_M), \\
q_2&\propto&\frac{1}{\sqrt2}(1-\varepsilon_M)(1-\Delta_M).
\label{piqi}
\end{eqnarray}

Therefore, up to the order of $O(\Delta_M)$ and $O(\varepsilon_M$),   we can simplify  Equations (\ref{m0}) and (\ref{m6}) as
\begin{eqnarray}
|M^0(t)\rangle &= &[g_+(t)-2\Delta_Mg_-(t)]|M_0\rangle+(1+2\varepsilon_M)g_-(t)|\bar{M}_0\rangle, \label{m0t}\\
|\bar{M}^0(t)\rangle&=&(1-2\varepsilon_M)g_-(t)|M_0\rangle+[g_+(t)+2\Delta_Mg_-(t)]|\bar{M}_0\rangle,
\label{m0bart}
\end{eqnarray}
where
\begin{eqnarray}
g_+\equiv\frac{1}{2}(e^{-i\lambda_1t}+e^{-i\lambda_2t}),  \\
g_-\equiv\frac{1}{2}(e^{-i\lambda_1t}-e^{-i\lambda_2t}).
\end{eqnarray}

If $CPT$ is conserved, then $H_{11}=H_{22}$, therefore  $\Delta_M=0$. Independent of the situation of $CPT$, if $CP$ is conserved, then $\varepsilon_M =0$. If $T$ is conserved, then $M_{12}^*/M_{12}=\Gamma_{12}^*/\Gamma_{12}$, therefore  $\Re\varepsilon_M=0$. If $CPT$ and $CP$ are both conserved, we have $T$ conservation. But $T$ conservation can also be valid without requiring the conservation of  $CPT$ and $CP$.  $T$ conservation together with $CP$ violation means $\Im\varepsilon_M\neq 0$.

\section{Correlated decays of entangled states  \label{amplitude} }

Suppose the initial state of two mesons $a$ and $b$ is the entangled  state of $C=\pm1$,
\begin{equation}
|\psi_\pm\rangle=\frac{1}{\sqrt{2}}[|M_0\rangle_a|\bar{M}_0\rangle_b\pm |\bar{M}_0\rangle_a|M_0\rangle_b].
\end{equation}
After time $t_a$, the final state of $a$ is $|f_a\rangle$, while after time $t_b$, the final state of $b$ is $|f_b\rangle$.

Define the decay amplitudes of $M_0$ and $\bar{M}_0$,
\begin{eqnarray}
g_1\equiv\langle f_a|M_0\rangle, \\
g_2\equiv\langle f_b|M_0\rangle, \\
h_1\equiv\langle f_a|\bar{M}_0\rangle, \\
h_2\equiv\langle f_b|\bar{M}_0\rangle.
\end{eqnarray}

The decay amplitude of the entangled meson pair can be obtained from (\ref{m0t}) and (\ref{m0bart}) as
\begin{equation}
A_{f_af_b}^\pm(t_a,t_b)=\frac{1}{\sqrt{2}}\big[\langle f_a|M_0(t_a)\rangle\langle f_b|\bar{M}_0(t_b)\rangle
\pm\langle f_a|\bar{M}_0(t_a)\rangle\langle f_b|M_0(t_b)\rangle\big].
\end{equation}
The decay rates for the entangled states are thus
\begin{eqnarray}
\Gamma_{f_af_b}^+(t_a, t_b)&=&|A_{f_af_b}^+(t_a, t_b)|^2   \nonumber \\
&=&\frac{1}{2}e^{-\Gamma_1(t_a+t_b)}|\alpha_1|^2+\frac{1}{2}e^{-\Gamma_2(t_a+t_b)}|\alpha_2|^2+\frac{1}{2}e^{-\Gamma_1t_a-\Gamma_2t_b}|\alpha_3|^2+\frac{1}{2}e^{-\Gamma_2t_a-\Gamma_1t_b}|\alpha_4|^2\nonumber \\
&&+e^{-\Gamma(t_a+t_b)}\Re[\alpha_1^*\alpha_2e^{-i\Delta m(t_a+t_b)}]+e^{-\Gamma_1t_a-\Gamma t_b}\Re[\alpha_1^*\alpha_3e^{-i\Delta mt_b}]\nonumber\\
&&+e^{-\Gamma t_a-\Gamma_1t_b}\Re[\alpha_1^*\alpha_4e^{-i\Delta mt_a}]+e^{-\Gamma t_a-\Gamma_2t_b}\Re[\alpha_2^*\alpha_3e^{i\Delta mt_a}]\nonumber\\
&&+e^{-\Gamma_2t_a-\Gamma t_b}\Re[\alpha_2^*\alpha_4e^{i\Delta mt_b}]+e^{-\Gamma(t_a+t_b)}\Re[\alpha_3^*\alpha_4e^{i\Delta m(t_b-t_a)}],
\label{gamma+}
\end{eqnarray}
\begin{eqnarray}
\Gamma_{f_af_b}^-(t_a, t_b)&=&|A_{f_af_b}^-(t_a, t_b)|^2  \nonumber  \\
&=&\frac{1}{2}e^{-\Gamma_1(t_a+t_b)}|\beta_1|^2+\frac{1}{2}e^{-\Gamma_2(t_a+t_b)}|\beta_2|^2+\frac{1}{2}e^{-\Gamma_1t_a-\Gamma_2t_b}|\beta_3|^2+\frac{1}{2}e^{-\Gamma_2t_a-\Gamma_1t_b}|\beta_4|^2\nonumber \\
&&+e^{-\Gamma(t_a+t_b)}\Re[\beta_1^*\beta_2e^{-i\Delta m(t_a+t_b)}]+e^{-\Gamma_1t_a-\Gamma t_b}\Re[\beta_1^*\beta_3e^{-i\Delta mt_b}]\nonumber\\
&&+e^{-\Gamma t_a-\Gamma_1t_b}\Re[\beta_1^*\beta_4e^{-i\Delta mt_a}]+e^{-\Gamma t_a-\Gamma_2t_b}\Re[\beta_2^*\beta_3e^{i\Delta mt_a}]\nonumber\\
&&+e^{-\Gamma_2t_a-\Gamma t_b}\Re[\beta_2^*\beta_4e^{i\Delta mt_b}]+e^{-\Gamma(t_a+t_b)}\Re[\beta_3^*\beta_4e^{i\Delta m(t_b-t_a)}],
\label{gamma-}
\end{eqnarray}
where
\begin{eqnarray*}
\Delta m&\equiv&m_2-m_1,\\
\Gamma&\equiv&\frac{\Gamma_1+\Gamma_2}{2},\\
\Delta\Gamma&\equiv&\Gamma_2-\Gamma_1,\\
\alpha_1&=&\frac{1}{2}(g_1+h_1)(g_2+h_2)+(\Delta_M+\varepsilon_M)(h_1h_2-g_1g_2)-(\Delta_M^2+\varepsilon_M^2)(g_1h_2+g_2h_1)\\
&&+2\Delta_M\varepsilon_M(g_1g_2+h_1h_2),\\
\alpha_2&=&-\frac{1}{2}(g_1-h_1)(g_2-h_2)+(\Delta_M-\varepsilon_M)(h_1h_2-g_1g_2)-(\Delta_M^2+\varepsilon_M^2)(g_1h_2+g_2h_1)\\
&&+2\Delta_M\varepsilon_M(g_1g_2+h_1h_2),\\
\alpha_3&=&\Delta_M(g_1+h_1)(g_2-h_2)+(\Delta_M^2+\varepsilon_M^2)(g_1h_2+g_2h_1)-2\Delta_M\varepsilon_M(g_1g_2+h_1h_2),\\
\alpha_4&=&\Delta_M(g_1-h_1)(g_2+h_2)+(\Delta_M^2+\varepsilon_M^2)(g_1h_2+g_2h_1)-2\Delta_M\varepsilon_M(g_1g_2+h_1h_2),\\
\beta_1&=&\beta_2=(\Delta_M^2-\varepsilon_M^2)(g_2h_1-g_1h_2),\\
\beta_3&=&\frac{1}{2}(g_1+h_1)(h_2-g_2)-\Delta_M(g_2h_1+g_1h_2)+\varepsilon_M(g_1g_2+h_1h_2)+(\Delta_M^2-\varepsilon_M^2)(g_1h_2-g_2h_1),\\
\beta_4&=&\frac{1}{2}(g_1-h_1)(g_2+h_2)+\Delta_M(g_2h_1+g_1h_2)-\varepsilon_M(g_1g_2+h_1h_2)+(\Delta_M^2-\varepsilon_M^2)(g_1h_2-g_2h_1).
\end{eqnarray*}

\section{Time-ordered integrated decay rates \label{integrated} }

Now consider two kinds of decay of an entangled state $|\psi_\pm\rangle$. One, whose integrated rate is denoted as $R^\pm_{ba}$, is that the decay of $b$ into $f_b$ precedes that of $a$ into $f_a$, i.e. $t_b \leq t_a$. The other, whose integrated rate is denoted as $R^\pm_{ab}$, is of the inverse time ordering, i.e. $t_b \geq t_a$. That is,
\begin{eqnarray}
R^\pm_{ba}&=&\int_0^\infty d{t_a}\int_0^{t_a}d{t_b}\Gamma_{f_af_b}^\pm({t_a},{t_b}), \\
R^\pm_{ab}&=&\int_0^\infty d{t_a}\int_{t_a}^\infty d{t_b}\Gamma_{f_af_b}^\pm({t_a},{t_b}).
\label{rsl1}
\end{eqnarray}

Making use of Equations (\ref{gamma+}) and (\ref{gamma-}), we obtain
\begin{eqnarray}
R_{ba}^+&=&\frac{|\alpha_1|^2}{4\Gamma_1^2}+\frac{|\alpha_2|^2}{4\Gamma_2^2}+\frac{|\alpha_3|^2}{4\Gamma_1\Gamma}+\frac{|\alpha_4|^2}{4\Gamma_2\Gamma}+\Re\big[\frac{\alpha_1^*\alpha_2}{2(\Gamma+i\Delta m)^2}\big]+\Re\big[\frac{\alpha_1^*\alpha_3}{\Gamma_1(\Gamma_1+\Gamma+i\Delta m)}\big]\nonumber\\
&&+\Re\big[\frac{\alpha_1^*\alpha_4}{(\Gamma+i\Delta m)(\Gamma_1+\Gamma+i\Delta m)}\big]+\Re\big[\frac{\alpha_2^*\alpha_3}{(\Gamma-i\Delta m)(\Gamma_2+\Gamma-i\Delta m)}\big] \nonumber \\
&&+\Re\big[\frac{\alpha_2^*\alpha_4}{\Gamma_2(\Gamma_2+\Gamma-i\Delta m)}\big]+\Re\big[\frac{\alpha_3^*\alpha_4}{2\Gamma(\Gamma+i\Delta m)}\big]\nonumber \\
&\simeq&A_0+A_1\Re\Delta_M+A_2\Im\Delta_M-A_3\Re\varepsilon_M-A_4\Im\varepsilon_M,\\
R_{ba}^-&=&\frac{|\beta_1|^2}{4\Gamma_1^2}+\frac{|\beta_2|^2}{4\Gamma_2^2}+\frac{|\beta_3|^2}{4\Gamma_1\Gamma}+\frac{|\beta_4|^2}{4\Gamma_2\Gamma}+\Re\big[\frac{\beta_1^*\beta_2}{2(\Gamma+i\Delta m)^2}\big]+\Re\big[\frac{\beta_1^*\beta_3}{\Gamma_1(\Gamma_1+\Gamma+i\Delta m)}\big]\nonumber\\
&&+\Re\big[\frac{\beta_1^*\beta_4}{(\Gamma+i\Delta m)(\Gamma_1+\Gamma+i\Delta m)}\big]+\Re\big[\frac{\beta_2^*\beta_3}{(\Gamma-i\Delta m)(\Gamma_2+\Gamma-i\Delta m)}\big] \nonumber \\
&&+\Re\big[\frac{\beta_2^*\beta_4}{\Gamma_2(\Gamma_2+\Gamma-i\Delta m)}\big]+\Re\big[\frac{\beta_3^*\beta_4}{2\Gamma(\Gamma+i\Delta m)}\big]\nonumber \\
&\simeq&B_0+B_1\Re\Delta_M+B_2\Im\Delta_M-B_3\Re\varepsilon_M-B_4\Im\varepsilon_M,\\
R_{ab}^+&=&\frac{|\alpha_1|^2}{4\Gamma_1^2}+\frac{|\alpha_2|^2}{4\Gamma_2^2}+\frac{|\alpha_3|^2}{4\Gamma_2\Gamma}+\frac{|\alpha_4|^2}{4\Gamma_1\Gamma}+\Re\big[\frac{\alpha_1^*\alpha_2}{2(\Gamma+i\Delta m)^2}\big]+\Re\big[\frac{\alpha_1^*\alpha_3}{(\Gamma+i\Delta m)(\Gamma_1+\Gamma+i\Delta m)}\big]\nonumber\\
&&+\Re\big[\frac{\alpha_1^*\alpha_4}{\Gamma_1(\Gamma_1+\Gamma+i\Delta m)}\big]+\Re\big[\frac{\alpha_2^*\alpha_3}{\Gamma_2(\Gamma_2+\Gamma-i\Delta m)}\big] \nonumber \\
&&+\Re\big[\frac{\alpha_2^*\alpha_4}{(\Gamma-i\Delta m)(\Gamma_2+\Gamma-i\Delta m)}\big]+\Re\big[\frac{\alpha_3^*\alpha_4}{2\Gamma(\Gamma-i\Delta m)}\big]\nonumber \\
&\simeq&C_0+C_1\Re\Delta_M+C_2\Im\Delta_M-C_3\Re\varepsilon_M-C_4\Im\varepsilon_M,\\
R_{ab}^-&=&\frac{|\beta_1|^2}{4\Gamma_1^2}+\frac{|\beta_2|^2}{4\Gamma_2^2}+\frac{|\beta_3|^2}{4\Gamma_2\Gamma}+\frac{|\beta_4|^2}{4\Gamma_1\Gamma}+\Re\big[\frac{\beta_1^*\beta_2}{2(\Gamma+i\Delta m)^2}\big]+\Re\big[\frac{\beta_1^*\beta_3}{(\Gamma+i\Delta m)(\Gamma_1+\Gamma+i\Delta m)}\big]\nonumber\\
&&+\Re\big[\frac{\beta_1^*\beta_4}{\Gamma_1(\Gamma_1+\Gamma+i\Delta m)}\big]+\Re\big[\frac{\beta_2^*\beta_3}{\Gamma_2(\Gamma_2+\Gamma-i\Delta m)}\big] \nonumber \\
&&+\Re\big[\frac{\beta_2^*\beta_4}{(\Gamma-i\Delta m)(\Gamma_2+\Gamma-i\Delta m)}\big]+\Re\big[\frac{\beta_3^*\beta_4}{2\Gamma(\Gamma-i\Delta m)}\big]\nonumber \\
&\simeq&D_0+D_1\Re\Delta_M+D_2\Im\Delta_M-D_3\Re\varepsilon_M-D_4\Im\varepsilon_M,
\label{rsl}
\end{eqnarray}
where   $\{A_i\},\{B_i\},\{C_i\}$ and $\{D_i\}(i=0,...,4)$ are given in the appendix.

Now we consider the asymmetries defined for these four time-ordered integrated decay rates,
\begin{eqnarray}
R_1\equiv\frac{R_{ba}^+-R_{ab}^+}{R_{ba}^++R_{ab}^+}, \\ R_2\equiv\frac{R_{ba}^--R_{ab}^-}{R_{ba}^-+R_{ab}^-}, \\
R_3\equiv\frac{R_{ba}^+-R_{ba}^-}{R_{ba}^++R_{ba}^-}, \\ R_4\equiv\frac{R_{ab}^+-R_{ab}^-}{R_{ab}^++R_{ab}^-},
\end{eqnarray}
which are calculated to be
\begin{eqnarray}
R_1&\simeq&\frac{A_1-C_1}{2A_0}\Re\Delta_M+\frac{A_2-C_2}{2A_0}\Im\Delta_M, \\
R_2&\simeq&\frac{B_0-D_0}{B_0+D_0}+\frac{2(B_1D_0-B_0D_1)}{(B_0+D_0)^2}
\Re\Delta_M+\frac{2(B_2D_0-B_0D_2)}{(B_0+D_0)^2}\Im\Delta_M \nonumber \\
&+&\frac{2(B_3D_0-B_0D_3)}{(B_0+D_0)^2}\Re\varepsilon_M+
\frac{2(B_4D_0-B_0D_4)}{(B_0+D_0)^2}\Im\varepsilon_M, \\
R_3&\simeq&\frac{A_0-B_0}{A_0+B_0}+\frac{2(A_1B_0-A_0B_1)}{(A_0+B_0)^2}
\Re\Delta_M+\frac{2(A_2B_0-A_0B_2)}{(A_0+B_0)^2}\Im\Delta_M \nonumber \\
&+&\frac{2(A_3B_0-A_0B_3)}{(A_0+B_0)^2}\Re\varepsilon_M+\frac{2(A_4B_0-A_0B_4)}{(A_0+B_0)^2}
\Im\varepsilon_M, \\
R_4&\simeq&\frac{C_0-D_0}{C_0+D_0}+\frac{2(C_1D_0-C_0D_1)}{(C_0+D_0)^2}\Re\Delta_M+
\frac{2(C_2D_0-C_0D_2)}{(C_0+D_0)^2}\Im\Delta_M \nonumber \\
&+&\frac{2(C_3D_0-C_0D_3)}{(C_0+D_0)^2}\Re\varepsilon_M+\frac{2(C_4D_0-C_0D_4)}{(C_0+D_0)^2}
\Im\varepsilon_M,
\label{ri}
\end{eqnarray}
which can be rewritten as
\begin{equation}
\left( \begin{array}{c}
R_1\\R_2-\frac{B_0-D_0}{B_0+D_0}\\R_3-\frac{A_0-B_0}{A_0+B_0}\\R_4-\frac{C_0-D_0}{C_0+D_0}
\end{array}\right)=K
\left( \begin{array}{c}
\Re\Delta_M\\ \Im\Delta_M\\ \Re\varepsilon_M\\ \Im\varepsilon_M
\end{array}\right),
\label{vio}
\end{equation}
with
\begin{equation}
K \equiv \left( \begin{array}{cccc}
\frac{A_1-C_1}{2A_0} & \frac{A_2-C_2}{2A_0} & 0 & 0\\
2\frac{B_1D_0-B_0D_1}{(B_0+D_0)^2} & 2\frac{B_2D_0-B_0D_2}{(B_0+D_0)^2} & 2\frac{B_3D_0-B_0D_3}{(B_0+D_0)^2} & 2\frac{B_4D_0-B_0D_4}{(B_0+D_0)^2}\\
2\frac{A_1B_0-A_0B_1}{(A_0+B_0)^2} & 2\frac{A_2B_0-A_0B_2}{(A_0+B_0)^2} & 2\frac{A_3B_0-A_0B_3}{(A_0+B_0)^2} & 2\frac{A_4B_0-A_0B_4}{(A_0+B_0)^2}\\
2\frac{C_1D_0-C_0D_1}{(C_0+D_0)^2} & 2\frac{C_2D_0-C_0D_2}{(C_0+D_0)^2} & 2\frac{C_3D_0-C_0D_3}{(C_0+D_0)^2} & 2\frac{C_4D_0-C_0D_4}{(C_0+D_0)^2}\\
\end{array}\right).
\end{equation}

Therefore,  $CP$ and $CPT$ violating  parameters are obtained as
\begin{equation}
\left( \begin{array}{c}
\Re\Delta_M\\ \Im\Delta_M\\ \Re\varepsilon_M\\ \Im\varepsilon_M
\end{array}\right)=K^{-1}
\left( \begin{array}{c}
R_1\\R_2-\frac{B_0-D_0}{B_0+D_0}\\R_3-\frac{A_0-B_0}{A_0+B_0}\\R_4-\frac{C_0-D_0}{C_0+D_0}
\end{array}\right),
\end{equation}
where, in terms of the Levi-Civita symbol $\epsilon_{ijkl}$,  $\Lambda_{ijk}\equiv(A_iB_0-A_0B_i)(B_jD_0-B_0D_j)(C_kD_0-C_0D_k)$ and  $\lambda_{ijk}\equiv(A_2-C_2)\epsilon_{i2jk}-(A_1-C_1)\epsilon_{1ijk}(i,j,k=1,2,3,4)$, the matrix elements of $K^{-1}$ are
\begin{eqnarray*}
(K^{-1})_{11}&=&-\frac{2A_0\epsilon_{1ijk}\Lambda_{ijk}}{\lambda_{ijk}\Lambda_{ijk}},\\
(K^{-1})_{12}&=&\frac{(A_2-C_2)(\Lambda_{324}-\Lambda_{423})(B_0+D_0)^2}{2(B_2D_0-B_0D_2)\lambda_{ijk}\Lambda_{ijk}},\\
(K^{-1})_{13}&=&\frac{(A_2-C_2)(\Lambda_{234}-\Lambda_{243})(A_0+B_0)^2}{2(A_2B_0-A_0B_2)\lambda_{ijk}\Lambda_{ijk}},\\
(K^{-1})_{14}&=&-\frac{(A_2-C_2)(\Lambda_{432}-\Lambda_{342})(C_0+D_0)^2}{2(C_2D_0-C_0D_2)\lambda_{ijk}\Lambda_{ijk}},\\
(K^{-1})_{21}&=&\frac{2A_0\epsilon_{i2jk}\Lambda_{ijk}}{\lambda_{ijk}\Lambda_{ijk}}, \\
(K^{-1})_{22}&=&\frac{(A_1-C_1)(\Lambda_{314}-\Lambda_{413})(B_0+D_0)^2}{2(B_1D_0-B_0D_1)\lambda_{ijk}\Lambda_{ijk}},\\
\end{eqnarray*}
\begin{eqnarray*}
(K^{-1})_{23}&=&-\frac{(A_1-C_1)(\Lambda_{134}-\Lambda_{143})(A_0+B_0)^2}{2(A_1B_0-A_0B_1)
\lambda_{ijk}\Lambda_{ijk}},\\
(K^{-1})_{24}&=&\frac{(A_1-C_1)(\Lambda_{431}-\Lambda_{341})(C_0+D_0)^2}{2(C_1D_0-C_0D_1)\lambda_{ijk}\Lambda_{ijk}}, \\
(K^{-1})_{31}&=&-\frac{2A_0\epsilon_{ij3k}\Lambda_{ijk}}{\lambda_{ijk}\Lambda_{ijk}}, \\
(K^{-1})_{32}&=&\frac{[(\Lambda_{432}-\Lambda_{234})(A_1-C_1)-(\Lambda_{431}-\Lambda_{134})(A_2-C_2)](B_0+D_0)^2}{2(B_3D_0-B_0D_3)\lambda_{ijk}\Lambda_{ijk}}, \\
(K^{-1})_{33}&=&\frac{[(\Lambda_{324}-\Lambda_{342})(A_1-C_1)-(\Lambda_{314}-\Lambda_{341})(A_2-C_2)](A_0+B_0)^2}{2(A_3B_0-A_0B_3)\lambda_{ijk}\Lambda_{ijk}}, \\
(K^{-1})_{34}&=&\frac{[(\Lambda_{243}-\Lambda_{423})(A_1-C_1)-(\Lambda_{143}-\Lambda_{413})(A_2-C_2)](C_0+D_0)^2}{2(C_3D_0-C_0D_3)\lambda_{ijk}\Lambda_{ijk}}, \\
(K^{-1})_{41}&=&\frac{2A_0\epsilon_{ijk4}\Lambda_{ijk}}{\lambda_{ijk}\Lambda_{ijk}}, \\
(K^{-1})_{42}&=&\frac{[(\Lambda_{243}-\Lambda_{342})(A_1-C_1)-(\Lambda_{143}-\Lambda_{341})(A_2-C_2)](B_0+D_0)^2}{2(B_4D_0-B_0D_4)\lambda_{ijk}\Lambda_{ijk}}, \\
(K^{-1})_{43}&=&\frac{[(\Lambda_{432}-\Lambda_{423})(A_1-C_1)-(\Lambda_{431}-\Lambda_{413})(A_2-C_2)](A_0+B_0)^2}{2(A_4B_0-A_0B_4)\lambda_{ijk}\Lambda_{ijk}}, \\
(K^{-1})_{44}&=&\frac{[(\Lambda_{324}-\Lambda_{234})(A_1-C_1)-(\Lambda_{314}-\Lambda_{134})(A_2-C_2)](C_0+D_0)^2}{2(C_3D_0-C_0D_3)\lambda_{ijk}\Lambda_{ijk}}.
\end{eqnarray*}

In general,  $CP$ and $CPT$ violating parameters can be obtained from the four asymmetries of the time-ordered integrated rates of the correlated decays of $CP=\pm1$ entangled states, using the above formulae. But in some cases, the four asymmetries may be independent of  $CP$ and $CPT$ violating parameters, as in one of the special cases discussed  below.

\section{Special Cases \label{special} }

We now consider the following two special cases~\cite{3}, for which the calculations are simplified a lot.

First, we consider the situation that $f_a=f$ while  $f_b=\bar{f}$, where $f$ and $\bar{f}$ are mutual CP conjugates and are decay products of $M^0$ and $\bar{M}^0$ respectively. For example, for $M^0=B^0$, $f=D^-D_S^+,D^-K^+,\pi^-D_S^+,\pi^-K^+$ while  $\bar{f}= D^+D_S^-,D^+K^-,\pi^+D_S^-,\pi^+K^-$, respectively.

In this case, the amplitudes satisfy $g_2=h_1=0$. Up to the order $O(\Delta_M)$ and $O(\varepsilon_M)$,  $\{\alpha_i\},\{\beta_i\}$ are given as
\begin{eqnarray}
\alpha_1&=&\alpha_2=\frac{1}{2}g_1h_2, \\
\alpha_3&=&-\alpha_4=-\Delta_M g_1 h_2, \\
\beta_1&=&\beta_2=0,\\
\beta_3&=&\big(\frac{1}{2}-\Delta_M\big)g_1h_2, \\
\beta_4&=&\big(\frac{1}{2}+\Delta_M\big)g_1h_2.
\end{eqnarray}
Then we obtain $A_{3,4}=B_{1,2,3,4}=C_{3,4}=D_{1,2,3,4}=0$. Using the expressions given in the appendix, Eq.(\ref{vio}) is simplified as
\begin{equation}
\left( \begin{array}{c}
R_1\\R_2-\frac{B_0-D_0}{B_0+D_0}\\R_3-\frac{A_0-B_0}{A_0+B_0}\\R_4-\frac{C_0-D_0}{C_0+D_0}
\end{array}\right)=
\left( \begin{array}{cccc}
\frac{A_1-C_1}{2A_0} & \frac{A_2-C_2}{2A_0} & 0 & 0\\
0 & 0 & 0 & 0\\
2\frac{A_1B_0}{(A_0+B_0)^2} & 2\frac{A_2B_0}{(A_0+B_0)^2} & 0 & 0\\
0 & 0 & 0 & 0\\
\end{array}\right)
\left( \begin{array}{c}
\Re\Delta_M\\\Im\Delta_M\\\Re\varepsilon_M\\\Im\varepsilon_M
\end{array}\right).
\end{equation}

In this case, we cannot obtain the $CP$ violating parameter from the four asymmetries, which are independent of  the former. Moreover, $R_2=\frac{B_0-D_0}{B_0+D_0}$ and $R_4=\frac{C_0-D_0}{C_0+D_0}$ are independent of both $CP$ and $CPT$ violating parameters, while $R_1$ and $R_3$  depend only on the $CPT$ violating parameter, but not on the $CP$ violating parameter.

Now we consider the situation that $f_a=f_b=f$, with  $f=\bar{f}$  being a $CP$ eigenstate. For example, $f_a=f_b=\pi^+\pi^-$,$\pi^0\pi^0$.
In this case, the decay amplitude satisfies $g_1=g_2$ and $h_1=h_2$. Hence the  $\{\alpha_i\},$ and  $\{\beta_i\}$ can be simplified as
\begin{eqnarray}
\alpha_1&=&\frac{1}{2}(g_1+h_1)^2+(\Delta_M+\varepsilon_M)(h_1^2-g_1^2)-2(\Delta_M^2+\varepsilon_M^2)g_1h_1+2\Delta_M\varepsilon_M(g_1^2+h_1^2), \\
\alpha_2&=&-\frac{1}{2}(g_1-h_1)^2+(\Delta_M-\varepsilon_M)(h_1^2-g_1^2)-2(\Delta_M^2+\varepsilon_M^2)g_1h_1+2\Delta_M\varepsilon_M(g_1^2+h_1^2), \\
\alpha_3&=&\alpha_4=\Delta_M(g_1^2-h_1^2)+2(\Delta_M^2+\varepsilon_M^2)g_1h_1-2\Delta_M\varepsilon_M(g_1^2+h_1^2),\\
\beta_1&=&\beta_2=0,\\
\beta_3&=&-\beta_4=\frac{1}{2}(g_1^2-h_1^2)-2\Delta_Mg_1h_1+\varepsilon_M(g_1^2+h_1^2)-2\Delta_Mg_1h_1+\varepsilon_M(g_1^2+h_1^2).
\label{absim}
\end{eqnarray}

Consequently $R_{ba}^-=R_{ab}^-=2|\varepsilon_M-\Delta_M|^2|g_1|^2\big(\frac{1}{\Gamma_1\Gamma_2}-\frac{1}{\Gamma^2+\Delta m^2}\big)$, which leads to $R_2=0$ exactly, no matter whether $CP$ or $CPT$ is violated.

If $CP$ is conserved, then $g_1=g_2=h_1=h_2$, hence  $\alpha_1=2g_1^2-2g_1^2(\Delta_M-\varepsilon_M)^2$, $\alpha_2=-\alpha_3=-\alpha_4=-2g_1^2(\Delta_M-\varepsilon_M)^2$, $\beta_3=-\beta_4=2g_1^2(\varepsilon_M-\Delta_M)$. Consequently, $R_{ba}^+=R_{ab}^+=\frac{|g_1|^4}{\Gamma_1^2}+O\big((\varepsilon_M-\Delta_M)^2\big)$. Then up to the order $O(\varepsilon_M)$ and $O(\Delta_M)$, we have $R_1=0$ and $R_3=R_4\approx1$.

\section{Discussions and summary \label{summary} }

Earlier, extensive discussions were made on  $|\psi_-\rangle$~\cite{buchanan,dambrosio}. However, most of these discussions concern the time-order asymmetry for integrated decay rates with a given time difference, rather than $R_2$, which is the asymmetry for the integrations over all time differences. Only for the case of $f_a=\pi^+\pi^-$ while $f_b=\pi^0\pi^0$, was $R_2$ calculated, with a nonzero value dependent on the $CP$ violating parameter~\cite{buchanan,dambrosio}. This is consistent with our results in Sec.~\ref{integrated}. Our results are in given in terms of the decay amplitudes defined for $|M_0\rangle$ and $|\bar{M}_0\rangle$, which are  related to the indirect $CP$ violating parameter $\epsilon'$.

In $\phi$ factory and in $\Upsilon (4s)$ resonance  in  B factories, the branch ratio of $C$-even meson pairs is negligibly small~\cite{dambrosio}, thus $R_{ab}^+$ and $R_{ba}^+$ cannot be measured there.  However, in $\Upsilon (5s)$ resonance  operated in  B factories such as CLEO and BELLE, the strong decay products include $B_s\bar{B}_s$, $B_s^*\bar{B}_s$ and   $B_s\bar{B}_s^*$.  As $B_s^* \rightarrow B_s\gamma$, the final states of $B_s^*\bar{B}_s$ and   $B_s\bar{B}_s^*$ are $B_s\bar{B}_s$ pair in $|\psi_+\rangle$. It turns out that the branch fraction of  $|\psi_-\rangle$ and $|\psi_+\rangle$ are $90\%$ and $10\%$, respectively \cite{soni,louvot}. Therefore, $\Gamma^{\pm}_{f_af_b}$ can both be measured once the correlated pairs are identified. In CPLEAR experiment on entangled kaons, the branch ratio between   $|\psi_+\rangle$ and $|\psi_-\rangle$ was as considerable as 0.037~\cite{cplear}. Moreover, we hope that in  future, neutral meson production can  start with an initial state of even orbital angular momentum so that  pairs in $|\psi_+\rangle$ dominate.

To summarize, we have presented a model-independent formalism of extracting rephase-invariant $CP$ and $CPT$ violating parameters from four asymmetries of time-ordered integrated rates of correlated decays of $C=\pm 1$ entangled states of neutral pseudoscalar mesons.   Asymmetries of time-ordered integrated rates are used because they are easy to be obtained from experimental data, without the necessity of measuring the exact times of decays.  We give the general  formulae applicable to all kinds of decay product. In the special case that $M^0$ and $\bar{M}^0$ decay to $CP$ conjugates, the asymmetries so defined cannot give information on the $CP$ violating parameter. Then one needs to use other observables, such as the time-ordered integrated rates themselves, or the asymmetries defined for specified differences of decay times. In future work, we will make more phenomenological studies and a more detailed comparison with the previous results~\cite{buchanan,dambrosio}.

\acknowledgments

This work was supported by the National Science Foundation of China (Grant No. 10875028).

\appendix

\section{Detailed expressions of $A_i$, $B_i$, $C_i$ and $D_i$}

\begin{eqnarray*}
A_0&=&\frac{|g_1+h_1|^2|g_2+h_2|^2}{8\Gamma_1^2}+\frac{|g_1-h_1|^2|g_2-h_2|^2}{8\Gamma_2^2}\\
&-&\frac{\Gamma^2-\Delta m^2}{4(\Gamma^2+\Delta m^2)^2}\big[(|g_1|^2-|h_1|^2)(|g_2|^2-|h_2|^2)-4\Im(g_1h_1^*)\Im(g_2h_2^*)\big],\\
A_1&=&\frac{1}{2\Gamma_1^2}\big[(|h_1h_2|^2-|g_1g_2|^2)-(|g_1|^2-|h_1|^2)\Re(g_2h_2^*)-(|g_2|^2-|h_2|^2)\Re(g_1h_1^*)\big]\\
&-&\frac{1}{2\Gamma_2^2}\big[|h_1h_2|^2-|g_1g_2|^2+(|g_1|^2-|h_1|^2)\Re(g_2h_2^*)+(|g_2|^2-|h_2|^2)\Re(g_1h_1^*)\big]\\
&-&\frac{\Gamma^2-\Delta m^2}{(\Gamma^2+\Delta m^2)^2}\big[(|g_1|^2-|h_1|^2)\Re(g_2h_2^*)+(|g_2|^2-|h_2|^2)\Re(g_1h_1^*)\big]\\
&+&\frac{\Gamma_1+\Gamma}{\Gamma_1[(\Gamma_1+\Gamma)^2+\Delta m^2]}|g_1+h_1|^2(|g_2|^2-|h_2|^2)+\frac{2\Delta m}{\Gamma_1[(\Gamma+\Gamma_1)^2+\Delta m^2]}|g_1+h_1|^2\Im(g_2h_2^*)\\
&-&\frac{\Gamma_2+\Gamma}{\Gamma_2[(\Gamma_2+\Gamma)^2+\Delta m^2]}|g_1-h_1|^2(|g_2|^2-|h_2|^2)-\frac{2\Delta m}{\Gamma_2[(\Gamma+\Gamma_2)^2+\Delta m^2]}|g_1-h_1|^2\Im(g_2h_2^*)\\
&+&\frac{\Gamma(\Gamma_1+\Gamma)-\Delta m^2}{(\Gamma^2+\Delta m^2)[(\Gamma_1+\Gamma)^2+\Delta m^2]}|g_2+h_2|^2(|g_1|^2-|h_1|^2)\\
&-&\frac{\Gamma(\Gamma_2+\Gamma)-\Delta m^2}{(\Gamma^2+\Delta m^2)[(\Gamma_2+\Gamma)^2+\Delta m^2]}|g_2-h_2|^2(|g_1|^2-|h_1|^2)\\
&+&\frac{2\Delta m(\Gamma_1+2\Gamma)}{(\Gamma^2+\Delta m^2)[(\Gamma_1+\Gamma)^2+\Delta m^2]}|g_2+h_2|^2\Im(g_1h_1^*)\\
&-&\frac{2\Delta m(\Gamma_2+2\Gamma)}{(\Gamma^2+\Delta m^2)[(\Gamma_2+\Gamma)^2+\Delta m^2]}|g_2-h_2|^2\Im(g_1h_1^*)\\
&-&\frac{4\Gamma\Delta m}{(\Gamma^2+\Delta m^2)^2}\Im(g_1h_1^*g_2h_2^*),\\
A_2&=&\frac{1}{2\Gamma_1^2}\big[(|g_1|^2+|h_1|^2)\Im(g_2h_2^*)+(|g_2|^2+|h_2|^2)\Im(g_1h_1^*)+2\Im(g_1h_1^*g_2h_2^*)\big]\\
&+&\frac{1}{2\Gamma_2^2}\big[(|g_1|^2+|h_1|^2)\Im(g_2h_2^*)+(|g_2|^2+|h_2|^2)\Im(g_1h_1^*)-2\Im(g_1h_1^*g_2h_2^*)\big]\\
&+&\frac{\Gamma^2-\Delta m^2}{(\Gamma^2+\Delta m^2)^2}\big[(|g_1|^2+|h_1|^2)\Im(g_2h_2^*)+(|g_2|^2+|h_2|^2)\Im(g_1h_1^*)\big]\\
&+&\frac{\Delta m(\Gamma_1+2\Gamma)}{(\Gamma^2+\Delta m^2)[(\Gamma_1+\Gamma)^2+\Delta m^2]}|g_2+h_2|^2(|g_1|^2-|h_1|^2)\\
&+&\frac{\Delta m(\Gamma_2+2\Gamma)}{(\Gamma^2+\Delta m^2)[(\Gamma_2+\Gamma)^2+\Delta m^2]}|g_2-h_2|^2(|g_1|^2-|h_1|^2)\\
&-&2\frac{\Gamma(\Gamma_1+\Gamma)-\Delta m^2}{(\Gamma^2+\Delta m^2)[(\Gamma_1+\Gamma)^2+\Delta m^2]}|g_2+h_2|^2\Im(g_1h_1^*)\\
&-&2\frac{\Gamma(\Gamma_2+\Gamma)-\Delta m^2}{(\Gamma^2+\Delta m^2)[(\Gamma_2+\Gamma)^2+\Delta m^2]}|g_2-h_2|^2\Im(g_1h_1^*)\\
\end{eqnarray*}
\begin{eqnarray*}
&+&\frac{\Delta m}{\Gamma_1[(\Gamma_1+\Gamma)^2+\Delta m^2]}|g_1+h_1|^2(|g_2|^2-|h_2|^2)\\
&+&\frac{\Delta m}{\Gamma_2[(\Gamma_2+\Gamma)^2+\Delta m^2]}|g_1-h_1|^2(|g_2|^2-|h_2|^2)\\
&-&\frac{2(\Gamma+\Gamma_1)}{\Gamma_1[(\Gamma_1+\Gamma)^2+\Delta m^2]}|g_1+h_1|^2\Im(g_2h_2^*)\\
&-&\frac{2(\Gamma_2+\Gamma)}{\Gamma_2[(\Gamma_2+\Gamma)^2+\Delta m^2]}|g_1-h_1|^2\Im(g_2h_2^*)\\
&-&\frac{2\Gamma\Delta m}{(\Gamma^2+\Delta m^2)^2}(|g_1g_2|^2-|h_1h_2|^2),\\
A_3&=&\frac{1}{2\Gamma_1}\big[|h_1h_2|^2-|g_1g_2|^2-(|g_1|^2-|h_1|^2)\Re(g_2h_2^*)-(|g_2|^2-|h_2|^2)\Re(g_1h_1^*)\big]\\
&+&\frac{1}{2\Gamma_2}\big[|h_1h_2|^2-|g_1g_2|^2+(|g_1|^2-|h_1|^2)\Re(g_2h_2^*)+(|g_2|^2-|h_2|^2)\Re(g_1h_1^*)\big]\\
&+&\frac{2\Gamma\Delta m}{(\Gamma^2+\Delta m^2)^2}\big[(|g_1|^2+|h_1|^2)\Im(g_2h_2^*)+(|g_2|^2+|h_2|^2)\Im(g_1h_1^*)\big]\\
&+&\frac{\Gamma^2-\Delta m^2}{(\Gamma^2+\Delta m^2)^2}(|g_1g_2|^2-|h_1h_2|^2),\\
A_4&=&\frac{1}{2\Gamma_1}\big[(|g_1|^2+|h_1|^2)\Im(g_2h_2^*)+(|g_2|^2+|h_2|^2)\Im(g_1h_1^*)+2\Im(g_1h_1^*g_2h_2^*)\big]\\
&-&\frac{1}{2\Gamma_2}\big[(|g_1|^2+|h_1|^2)\Im(g_2h_2^*)+(|g_2|^2+|h_2|^2)\Im(g_1h_1^*)-2\Im(g_1h_1^*g_2h_2^*)\big]\\
&+&\frac{2\Gamma\Delta m}{(\Gamma^2+\Delta m^2)^2}\big[(|g_1|^2-|h_1|^2)\Re(g_2h_2^*)+(|g_2|^2-|h_2|^2)\Re(g_1h_1^*)\big]\\
&-&2\frac{\Gamma^2-\Delta m^2}{(\Gamma^2+\Delta m^2)^2}\Im(g_1h_1^*g_2h_2^*),\\
B_0&=&\frac{1}{8\Gamma\Gamma_1}|g_1+h_1|^2|g_2-h_2|^2+\frac{1}{8\Gamma\Gamma_2}|g_1-h_1|^2|g_2+h_2|^2\\
&-&\frac{1}{4(\Gamma^2+\Delta m^2)}\big[(|g_1|^2-|h_1|^2)(|g_2|^2-|h_2|^2)+4\Im(g_1h_1^*)\Im(g_2h_2^*)\big]\\
&+&\frac{\Delta m}{2\Gamma(\Gamma^2+\Delta m^2)}\big[(|g_1|^2-|h_1|^2)\Im(g_2h_2^*)-(|g_2|^2-|h_2|^2)\Im(g_1h_1^*)\big],\\
B_1&=&\frac{1}{2\Gamma\Gamma_2}\big[|g_1h_2|^2-|g_2h_1|^2+(|g_1|^2-|h_1|^2)\Re(g_2h_2^*)+(|g_2|^2-|h_2|^2)\Re(g_1h_1^*)\big]\\
&-&\frac{1}{2\Gamma\Gamma_1}\big[|g_1h_2|^2-|g_2h_1|^2-(|g_1|^2-|h_1|^2)\Re(g_2h_2^*)-(|g_2|^2-|h_2|^2)\Re(g_1h_1^*)\big]\\
&-&\frac{1}{\Gamma^2+\Delta m^2}\big[(|g_1|^2-|h_1|^2)\Re(g_2h_2^*)+(|g_2|^2-|h_2|^2)\Re(g_1h_1^*)\big]-\frac{2\Delta m}{\Gamma(\Gamma^2+\Delta m^2)}\Im(g_1h_1^*g_2h_2^*),\\
B_2&=&\frac{1}{2\Gamma\Gamma_2}\big[2\Im(g_1h_1^*g_2^*h_2)+(|g_1|^2+|h_1|^2)\Im(g_2h_2^*)+(|g_2|^2+|h_2|^2)\Im(g_1h_1^*)\big]\\
&-&\frac{1}{2\Gamma\Gamma_1}\big[-2\Im(g_1h_1^*g_2^*h_2)+(|h_1|^2+|g_1|^2)\Im(g_2h_2^*)+(|h_2|^2+|g_2|^2)\Im(g_1h_1^*)\big]\\
\end{eqnarray*}
\begin{eqnarray*}
&-&\frac{1}{\Gamma^2+\Delta m^2}\big[(|g_1|^2+|h_1|^2)\Im(g_2h_2^*)+(|g_2|^2+|h_2|^2)\Im(g_1h_1^*)\big]\\
&+&\frac{\Delta m}{\Gamma(\Gamma^2+\Delta m^2)}(|g_1h_2|^2-|g_2h_1|^2),\\
B_3&=&\frac{1}{2\Gamma\Gamma_2}\big[|g_1g_2|^2-|h_1h_2|^2+(|g_1|^2-|h_1|^2)\Re(g_2h_2^*)-(|g_2|^2-|h_2|^2)\Re(g_2h_2^*)\big]\\
&+&\frac{1}{2\Gamma\Gamma_1}\big[|h_1h_2|^2-|g_1g_2|^2+(|g_1|^2-|h_1|^2)\Re(g_2h_2^*)-(|g_2|^2-|h_2|^2)\Re(g_2h_2^*)\big]\\
&-&\frac{\Delta m}{\Gamma(\Gamma^2+\Delta m^2)}\big[(|g_1|^2+|h_1|^2)\Im(g_2h_2^*)-(|g_2|^2+|h_2|^2)\Im(g_1h_1^*)\big]\\
&+&\frac{1}{\Gamma^2+\Delta m^2}(|g_1g_2|^2-|h_1h_2|^2),\\
B_4&=&\frac{1}{2\Gamma\Gamma_2}\big[2\Im(g_1h_1^*g_2h_2^*)+(|g_2|^2+|h_2|^2)\Im(g_1h_1^*)-(|g_1|^2+|h_1|^2)\Im(g_2h_2^*)\big]\\
&-&\frac{1}{2\Gamma\Gamma_1}\big[2\Im(g_1h_1^*g_2h_2^*)+(|g_1|^2+|h_1|^2)\Im(g_2h_2^*)-(|g_2|^2+|h_2|^2)\Im(g_1h_1^*)\big]\\
&-&\frac{\Delta m}{\Gamma(\Gamma^2+\Delta m^2)}\big[(|g_1|^2-|h_1|^2)\Re(g_2h_2^*)-(|g_2|^2-|h_2|^2)\Re(g_1h_1^*)\big]\\
&+&\frac{2}{\Gamma^2+\Delta m^2}\Im(g_1h_1^*g_2h_2^*),\\
C_0&=&A_0,C_3=A_3,C_4=A_4,\\
C_1&=&\frac{1}{2\Gamma_1^2}\big[|h_1h_2|^2-|g_1g_2|^2-(|g_1|^2-|h_1|^2)\Re(g_2h_2^*)-(|g_2|^2-|h_2|^2)\Re(g_1h_1^*)\big]\\
&-&\frac{1}{2\Gamma_2^2}\big[|h_1h_2|^2-|g_1g_2|^2+(|g_1|^2-|h_1|^2)\Re(g_2h_2^*)+(|g_2|^2-|h_2|^2)\Re(g_1h_1^*)\big]\\
&-&\frac{\Gamma^2-\Delta m^2}{(\Gamma^2+\Delta m^2)^2}\big[(|g_1|^2-|h_1|^2)\Re(g_2h_2^*)+(|g_2|^2-|h_2|^2)\Re(g_1h_1^*)\big]\\
&+&\frac{\Gamma_1+\Gamma}{\Gamma_1[(\Gamma_1+\Gamma)^2+\Delta m^2]}|g_2+h_2|^2(|g_1|^2-|h_1|^2)+\frac{2\Delta m}{\Gamma_1[(\Gamma+\Gamma_1)^2+\Delta m^2]}|g_2+h_2|^2\Im(g_1h_1^*)\\
&-&\frac{\Gamma_2+\Gamma}{\Gamma_2[(\Gamma_2+\Gamma)^2+\Delta m^2]}|g_2-h_2|^2(|g_1|^2-|h_1|^2)-\frac{2\Delta m}{\Gamma_2[(\Gamma+\Gamma_2)^2+\Delta m^2]}|g_2-h_2|^2\Im(g_1h_1^*)\\
&+&\frac{\Gamma(\Gamma_1+\Gamma)-\Delta m^2}{(\Gamma^2+\Delta m^2)[(\Gamma_1+\Gamma)^2+\Delta m^2]}|g_1+h_1|^2(|g_2|^2-|h_2|^2)\\
&-&\frac{\Gamma(\Gamma_2+\Gamma)-\Delta m^2}{(\Gamma^2+\Delta m^2)[(\Gamma_2+\Gamma)^2+\Delta m^2]}|g_1-h_1|^2(|g_2|^2-|h_2|^2)\\
&-&\frac{2\Delta m(\Gamma_2+2\Gamma)}{(\Gamma^2+\Delta m^2)[(\Gamma_2+\Gamma)^2+\Delta m^2]}|g_1-h_1|^2\Im(g_2h_2^*)\\
&+&\frac{2\Delta m(\Gamma_1+2\Gamma)}{(\Gamma^2+\Delta m^2)[(\Gamma_1+\Gamma)^2+\Delta m^2]}|g_1+h_1|^2\Im(g_2h_2^*)\\
&-&\frac{4\Gamma\Delta m}{(\Gamma^2+\Delta m^2)^2}\Im(g_1h_1^*g_2h_2^*),\\
\end{eqnarray*}
\begin{eqnarray*}
C_2&=&\frac{1}{2\Gamma_1^2}\big[(|g_1|^2+|h_1|^2)\Im(g_2h_2^*)+(|g_2|^2+|h_2|^2)\Im(g_1h_1^*)+2\Im(g_1h_1^*g_2h_2^*)\big]\\
&+&\frac{1}{2\Gamma_2^2}\big[(|g_1|^2+|h_1|^2)\Im(g_2h_2^*)+(|g_2|^2+|h_2|^2)\Im(g_1h_1^*)-2\Im(g_1h_1^*g_2h_2^*)\big]\\
&+&\frac{\Gamma^2-\Delta m^2}{(\Gamma^2+\Delta m^2)^2}\big[(|g_1|^2+|h_1|^2)\Im(g_2h_2^*)+(|g_2|^2+|h_2|^2)\Im(g_1h_1^*)\big]\\
&-&\frac{2(\Gamma+\Gamma_1)}{\Gamma_1[(\Gamma_1+\Gamma)^2+\Delta m^2]}|g_2+h_2|^2\Im(g_1h_1^*)-\frac{2(\Gamma_2+\Gamma)}{\Gamma_2[(\Gamma_2+\Gamma)^2+\Delta m^2]}|g_2-h_2|^2\Im(g_1h_1^*)\\
&+&\frac{\Delta m(\Gamma_1+2\Gamma)}{(\Gamma^2+\Delta m^2)[(\Gamma_1+\Gamma)^2+\Delta m^2]}|g_1+h_1|^2(|g_2|^2-|h_2|^2)-\frac{2\Gamma\Delta m}{(\Gamma^2+\Delta m^2)^2}(|g_1g_2|^2-|h_1h_2|^2)\\
&+&\frac{\Delta m(\Gamma_2+2\Gamma)}{(\Gamma^2+\Delta m^2)[(\Gamma_2+\Gamma)^2+\Delta m^2]}|g_1-h_1|^2(|g_2|^2-|h_2|^2)\\
&-&2\frac{\Gamma(\Gamma_1+\Gamma)-\Delta m^2}{(\Gamma^2+\Delta m^2)[(\Gamma_1+\Gamma)^2+\Delta m^2]}|g_1+h_1|^2\Im(g_2h_2^*)\\
&-&2\frac{\Gamma(\Gamma_2+\Gamma)-\Delta m^2}{(\Gamma^2+\Delta m^2)[(\Gamma_2+\Gamma)^2+\Delta m^2]}|g_1-h_1|^2\Im(g_2h_2^*)\\
&+&\frac{\Delta m}{\Gamma_2[(\Gamma_2+\Gamma)^2+\Delta m^2]}|g_2-h_2|^2(|g_1|^2-|h_1|^2)\\
&+&\frac{\Delta m}{\Gamma_1[(\Gamma_1+\Gamma)^2+\Delta m^2]}|g_2+h_2|^2(|g_1|^2-|h_1|^2),\\
D_0&=&\frac{1}{8\Gamma\Gamma_1}|g_1-h_1|^2|g_2+h_2|^2+\frac{1}{8\Gamma\Gamma_2}|g_1+h_1|^2|g_2-h_2|^2\\
&-&\frac{1}{4(\Gamma^2+\Delta m^2)}\big[(|g_1|^2-|h_1|^2)(|g_2|^2-|h_2|^2)+4\Im(g_1h_1^*)\Im(g_2h_2^*)\big]\\
&-&\frac{\Delta m}{2\Gamma(\Gamma^2+\Delta m^2)}\big[(|g_1|^2-|h_1|^2)\Im(g_2h_2^*)-(|g_2|^2-|h_2|^2)\Im(g_1h_1^*)\big],\\
D_1&=&\frac{1}{2\Gamma\Gamma_1}\big[|g_1h_2|^2-|g_2h_1|^2+(|g_1|^2-|h_1|^2)\Re(g_2h_2^*)+(|g_2|^2-|h_2|^2)\Re(g_1h_1^*)\big]\\
&-&\frac{1}{2\Gamma\Gamma_2}\big[|g_1h_2|^2-|g_2h_1|^2-(|g_1|^2-|h_1|^2)\Re(g_2h_2^*)-(|g_2|^2-|h_2|^2)\Re(g_1h_1^*)\big]\\
&-&\frac{1}{\Gamma^2+\Delta m^2}\big[(|g_1|^2-|h_1|^2)\Re(g_2h_2^*)+(|g_2|^2-|h_2|^2)\Re(g_1h_1^*)\big]\\
&-&\frac{2\Delta m}{\Gamma(\Gamma^2+\Delta m^2)}\Im(g_1h_1^*g_2h_2^*),\\
D_2&=&\frac{1}{2\Gamma\Gamma_1}\big[2\Im(g_1h_1^*g_2^*h_2)+(|h_1|^2+|g_1|^2)\Im(g_2h_2^*)+(|h_2|^2+|g_2|^2)\Im(g_1h_1^*)\big]\\
&-&\frac{1}{2\Gamma\Gamma_2}\big[2\Im(g_1h_1^*g_2^*h_2)-(|h_1|^2+|g_1|^2)\Im(g_2h_2^*)-(|h_2|^2+|g_2|^2)\Im(g_1h_1^*)\big]\\
&-&\frac{1}{\Gamma^2+\Delta m^2}\big[(|g_1|^2+|h_1|^2)\Im(g_2h_2^*)+(|g_2|^2+|h_2|^2)\Im(g_1h_1^*)\big]\\
&-&\frac{\Delta m}{\Gamma(\Gamma^2+\Delta m^2)}(|g_1h_2|^2-|g_2h_1|^2),\\
\end{eqnarray*}
\begin{eqnarray*}
D_3&=&\frac{1}{2\Gamma\Gamma_1}\big[|g_1g_2|^2-|h_1h_2|^2+(|g_1|^2-|h_1|^2)\Re(g_2h_2^*)-(|g_2|^2-|h_2|^2)\Re(g_1h_1^*)\big]\\
&-&\frac{1}{2\Gamma\Gamma_2}\big[|g_1g_2|^2-|h_1h_2|^2-(|g_1|^2-|h_1|^2)\Re(g_2h_2^*)+(|g_2|^2-|h_2|^2)\Re(g_1h_1^*)\big]\\
&+&\frac{\Delta m}{\Gamma(\Gamma^2+\Delta m^2)}\big[(|g_1|^2+|h_1|^2)\Im(g_2h_2^*)-(|g_2|^2+|h_2|^2)\Im(g_1h_1^*)\big]\\
&+&\frac{1}{\Gamma^2+\Delta m^2}(|g_1g_2|^2-|h_1h_2|^2),\\
D_4&=&\frac{1}{2\Gamma\Gamma_1}\big[2\Im(g_1h_1^*g_2^*h_2)-(|h_1|^2+|g_1|^2)\Im(g_2h_2^*)+(|h_2|^2+|g_2|^2)\Im(g_1h_1^*)\big]\\
&-&\frac{1}{2\Gamma\Gamma_2}\big[2\Im(g_1h_1^*g_2^*h_2)+(|h_1|^2+|g_1|^2)\Im(g_2h_2^*)-(|h_2|^2+|g_2|^2)\Im(g_1h_1^*)\big]\\
&+&\frac{\Delta m}{\Gamma(\Gamma^2+\Delta m^2)}\big[(|g_1|^2-|h_1|^2)\Re(g_2h_2^*)-(|g_2|^2-|h_2|^2)\Re(g_1h_1^*)\big]\\
&+&\frac{2}{\Gamma^2+\Delta m^2}\Im(g_1h_1^*g_2h_2^*).
\end{eqnarray*}

\end{document}